\documentclass[%
 reprint,
 amsmath,amssymb,
 aps,
floatfix,
unsortedaddress
]{revtex4-1}

\usepackage[T1]{fontenc}
\usepackage{graphicx}
\usepackage{dcolumn}
\usepackage{bm}
\usepackage{xcolor}

\begin{document}

\preprint{APS/123-QED}

\title{Entanglement generation in a quantum network with finite quantum memory lifetime}

\author{Vyacheslav Semenenko}
\affiliation{Department of Electrical Engineering, University at Buffalo, Buffalo,
NY 14260, USA}
\author{Xuedong Hu}
\affiliation{Department of Physics, University at Buffalo, Buffalo,
NY 14260, USA}
\author{Eden Figueroa}
\affiliation{Department of Physics and Astronomy, Stony Brook University, Stony Brook,
NY 11794, USA}
\author{Vasili Perebeinos}
\email{vasilipe@buffalo.edu}
\affiliation{Department of Electrical Engineering, University at Buffalo, Buffalo,
NY 14260, USA}

\date{\today}

\begin{abstract}
We simulate entanglement sharing between two end-nodes of a linear chain quantum network using SeQUeNCe, an open-source simulation package for quantum networks. Our focus is on the rate of entanglement generation between the end-nodes with many repeaters with a finite quantum memory lifetime. Numerical and analytical simulations show limits of connection performance for a given number of repeaters involved, memory lifetimes, the distance between the end-nodes, and an entanglement management protocol. Our findings demonstrate that the performance of quantum connection depends highly on the entanglement management protocol, which schedules entanglement generation and swapping, resulting in the final end-to-end entanglement. 
\end{abstract}

\maketitle


\section{Introduction}

In recent years, many research efforts have focused on developing quantum  communication networks and sharing quantum entanglement among spatially separated qubits~\cite{Sangouard2011,Bao2012,Chen2007,Jin2015,Sun2017,Leent2020,Dideriksen2021,Pompili2021,Duan2001, Yang2013EntanglementGeneration}.  Small linear quantum networks (with maximum distance in a pair of nodes up to about 100 km) have already been demonstrated~\cite{Wang2021, Eden2021}. Theoretical development of network protocols has reached its third generation. The generations are classified by methods adopted to suppress loss and operation errors by either heralded entanglement generation or quantum error correction~\cite{lukin2016generations}. The first two generations are based on heralded entanglement generation, requiring signaling back to communicating nodes about the generation status. The third generation of the quantum network is free from this obligation that significantly limits the network throughput. However, the third generation networks make severe demands on the fidelity of quantum gates that are now incompatible with today's hardware's characteristics~\cite{flamini2018review,lukin2016generations}. 

Currently, available quantum hardware components are still far from allowing the realization of a fully functional first-generation network for long-distance end-to-end quantum communication~\cite{wehner2018vision}.  Key bottlenecks include limited coherence time of qubits and photon loss in the quantum channel medium between the end-nodes. While the latter issue can, in principle, be overcome by using quantum repeaters~\cite{Chen2007,Bao2012, mmm2007prl, rate-loss2015pra, Kok2017,simon2020nearterm, vardoyan2021stoch}, including multimode solutions with multiplexing~\cite{multiplex3,simon2014multiplexing2,Pu2017,Dam2017,Forbes2020,Lipka2021,multiplexing1}, the former has been a persistent impediment to progress in quantum communication in particular and quantum information technologies in general. The multi-path routing~\cite{Laurenza2017,Pant2019}, involving multiple paths for routing entanglement between a pair of end-users, can enable a long-distance entanglement generation rate with a higher rate than what is possible with a linear repeater chain~\cite{Pirandola2019}.

Quantum memories are essential  ingredients for a quantum repeater~\cite{Jing2019,Boch2018}.  Their coherent lifetimes are crucial for creating and maintaining high fidelity entanglement~\cite{Yu2020,Yang2016,Bhaskar2020} and are dependent on the materials platform. While typical spin qubits coherence time of a few milliseconds up to a second in silicon~\cite{Hanson2007,Pla2013} and diamond~\cite{Balasubramanian2009,Awschalom2018,Abobeih2018} are comparable to an average ping time of about 0.1 milliseconds in classical networks, the trapped ions demonstrate memory lifetimes from several minutes to hours~\cite{Langer2005,Kotler2014,Wang2021b}. However, the frequency conversion efficiency to telecommunication wavelength ($\sim$1560 $\mu$m) remains low~\cite{Kwiat1995,Boyd,Schneeloch2019}. Currently, there is intense research for quantum memory development based on novel two-dimensional materials~\cite{Zhao2021}, rare Earth ion-doped optical fibers~\cite{Saglamyurek2015,Lago-Rivera2021}, and quantum dots~\cite{Kim2017}  which would operate at the telecommunication wavelength of the quantum memory developments, thus obviating the need for the frequency conversion step. The variety of materials platforms~\cite{interconnects2021,Guha2021} for the quantum network components makes the design of quantum networks a challenging engineering task. Preliminary simulations are necessary for designing optical-fiber classical connections specifically for quantum networks~\cite{Dahlberg_2018,sequence2019qkd,Matsuo2019,Ghalaii2020,sequence2020arxiv,coopmans2021netsquid,9465750}. This kind of simulation allows a designer to construct the most tolerant protocol to account for classical network ping, single-photon traveling time fluctuations, and other parameter imperfections in the quantum network hardware and evaluate if the existing hardware satisfies the error-robustness requirements.

One of the essential resources for a quantum network is the entanglement between any two network nodes, which would allow the transfer of unknown states between the nodes via quantum teleportation. Many protocols have been suggested to generate and distribute such remote entanglement~\cite{Maruyama2007EntanglementTransfer, Zhu2015EntanglementDistribution, Qin2018LightMatterInt, Gneiting2019EntanglementTransport}, with photonic networks being the most mature and technologically realistic approach. For example, one family of protocols~\cite{marcikic2002swapping,azuma2015qmemfree-theory,li2019qmemfree-exp} aims to exclude troubles associated with imperfection of quantum memories in intermediate nodes. However, this requires that all the intermediate links between the end-nodes succeed simultaneously. Such an approach reduces the impact of qubit decoherence, but the need for synchronicity can be demanding. Other protocols rely on qubits to preserve an entangled state for a long time until all the links are successfully established between the end-nodes. Such an approach reduces the number of attempts but requires the qubits to stay in the entangled states with high fidelity for long times~\cite{zoller1999purif,hartmann2007memerr,Harty2014,vardoyan2021stoch}.

Here we analyze the entanglement generation capabilities between the end-nodes in a simple linear network using SeQEeNCe quantum network simulator~\citep{sequence2019qkd,sequence2020arxiv}. Although several codes are available for simulating quantum networks with similar capabilities, we decided to use SeQUeNCe because it is easy to modify the code for non-developers. We examine the two protocols discussed above in detail and apply them to the generation of elementary links between adjacent nodes.  Our results for both scenarios set the requirements on the number of repeaters and memory lifetime for different distances between the end-nodes.


\begin{figure}[htp]
\centering{\includegraphics[width=0.99\columnwidth]{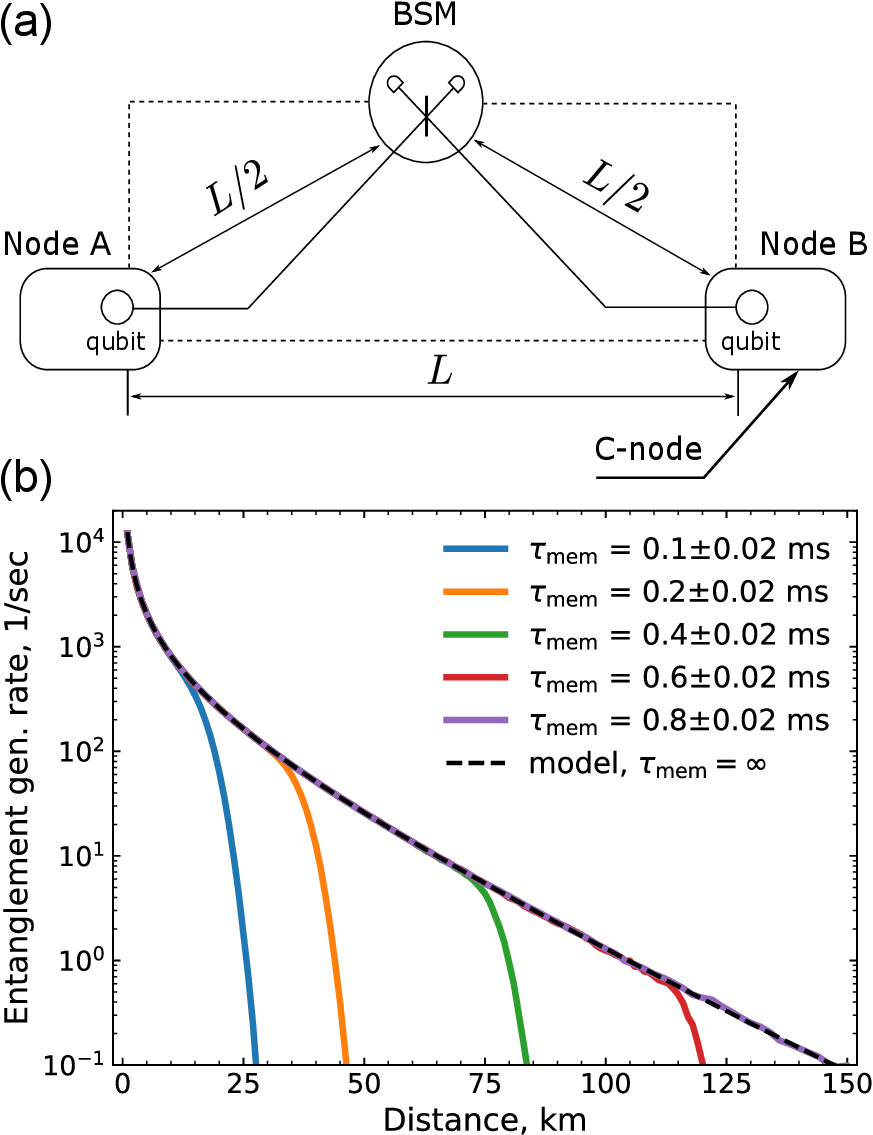}}
\caption{(a) Schematics of an elementary quantum link implementing Barrett-Kok protocol for entanglement generation. Solid lines show optical fiber connections for quantum information channels, and dashed lines designate connections for classical information.  (b) Simulated entanglement generation rate for different memory lifetimes of the qubits using SeQUeNCe. The simulations assume the light velocity in the quantum channel and signal velocity in the classical channel  $v=2\times10^{5}$
km/s, photon attenuation rate in quantum channels $\alpha=0.2$ dB/km,
memories efficiency $\mathcal{E}_{\mathrm{m}}=90\%$ and detectors
efficiency $\mathcal{E}_{\mathrm{d}}=80\%$. The solid line shows the results from the analytical model given by Eq.~\eqref{eq:rate-no-reps}.}
\label{fig:elementary}
\end{figure}

\section{ Generation of the elementary quantum link}
An elementary link between two quantum network nodes is established when quantum entanglement is created between a pair of qubits belonging to the nodes. The entanglement, once created, is a resource used for quantum information transmission from one node to the other. Once a quantum state from the sending node is transmitted, the entanglement is destroyed. An essential parameter characterizing the performance of a given quantum connection is the maximum rate at which entanglement is generated.

Many protocols have been proposed to generate entanglement between two qubits.  Here we choose to analyze the Barrett-Kok protocol’s performance~\cite{barrett-kok2005pra} for entanglement generation between two nodes with and without intermediate nodes (quantum repeaters). The Barrett-Kok protocol is robust against losses, and it has been implemented experimentally in several studies~\cite{Hensen2015,Englund2020arxiv}.
Our analysis can be generalized to other protocols straightforwardly. An elementary link in the Barrett-Kok protocol is
shown in Fig.~\ref{fig:elementary}.  The nodes are located at a distance $L$ from each other, and the Bell-state measuring (BSM) station is located right in the middle (in terms of the length of optical communication fiber).  We assume that synchronization and scheduling instructions start with basic operations, such as single-photon emission. The instructions are obtained from a single node, called C-node (controlling node), one of the nodes forming a chain and participating in the end-to-end quantum connection. In Fig.~\ref{fig:elementary}a, there are no intermediate nodes between the end-nodes, so an end-node is chosen as a control node. It is most efficient to assign C-node duties to the middle node in the chain, so the maximum ping to the other nodes is minimal. The presence of C-node makes the system hierarchic, i.e., managed from the single-center that allows us to focus on the analysis of the performance of the quantum network and frees us from developing more complex communication protocols specific to peer-to-peer networks~\cite{p2p-book}.

\begin{figure*}[htp]
\centering{\includegraphics[width=0.8\textwidth]{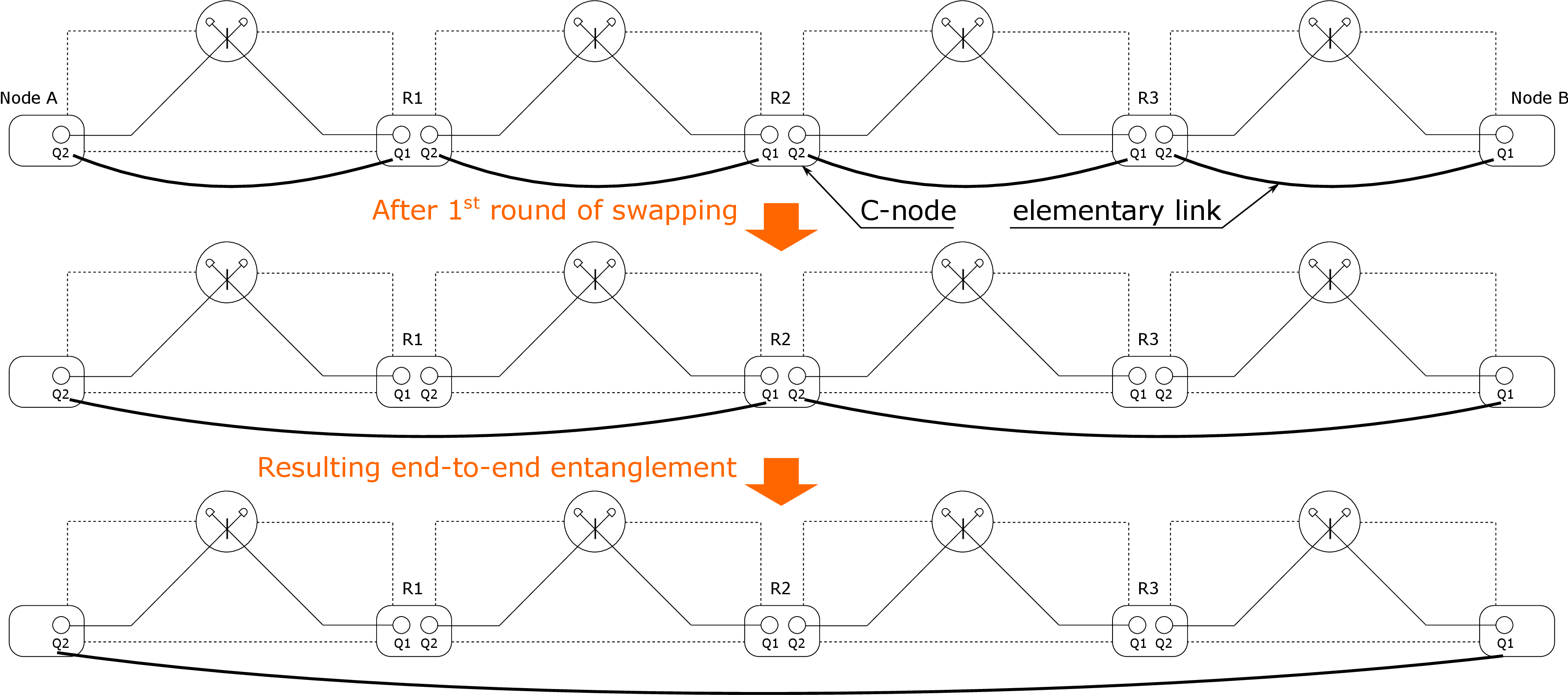}}
\caption{Schematics of entanglement swapping to establish entanglement between the end-nodes connected via three quantum repeaters ($r=3$). The entanglement swapping iterations $\tilde{n}=2$ are shown on the two bottom panels.}
\label{fig:rep}
\end{figure*}

The entanglement generation is organized in the following stages. At time $t=0$, one of the nodes that acts as the C-node sends a message to another node to start the Barrett-Kok protocol. At timestamp $t=L/v$, where $v$ is the photon velocity in optical fiber, the protocol start time is received. Once the message is received, immediately (assuming no delay) the nodes excite their qubits so that the qubit would then be entangled with the photon in the optical cavity it resides in. Photon emission from the cavity happens with the probability $\mathcal{E}_{\mathrm{m}}$ called memory efficiency. At timestamp $t=2.5L/v$ the leaked photons should reach the BSM station, where the measurement takes place. For simplicity, we assume that the signal speed in the classical channel is $v$, the same as in the quantum channel, and the lengths of the corresponding quantum and classical optical channels are the same. Typically, dark fiber infrastructure for secure communication is installed parallel with the standard optical fibers. Therefore, it is natural to assume that optical paths for the classical and quantum signals are the same. Thus, the result of measurements in BSM is received at timestamp $t=3L/v$. Additionally, BSM measurements in the Barrett-Kok protocol are done two times successively, and once results of the second round of measurements are obtained, both nodes know the final result of the entanglement generation attempt at timestamp $t=4L/v$. If the entanglement is successfully generated, both nodes immediately use it according to their program (e.g., for quantum teleportation or simultaneous measurement in different bases as required in the BB84 protocol~\cite{bennett1984,bennett1984}).  We increase the counter of generated entanglements by one. Neglecting delays in the measurements and classical messages formation, we assume that at a timestamp of $t=4L/v$ the described steps above are repeated from the beginning.

After escaping their cavities, photons can be lost in the fiber due to attenuation,  which we model by the decay probability
 $e^{-\gamma L/2}$,
$\gamma=\alpha/10\cdot\log10$, where $\alpha$ is the attenuation constant
(given in dB per length) and $L/2$ is the fiber length, as shown in Fig.~\ref{fig:elementary}a. The BSM station contains four single-photon detectors,
each detecting a photon with probability $\mathcal{E}_{\mathrm{d}}$ called detector efficiency.
If both photons are detected, two out of four Bell-states can be measured by this scheme.
Ideally, the success rate for the Barrett-Kok protocol for entanglement swapping with BSM is $\mathcal{E}_{\mathrm{b}}=$50\%~\citep{barrett-kok2005pra}.  By combining probabilities for the length-dependent losses, two detectors, two quantum memories, and one BSM swapping, we arrive at  the success probability for establishing entanglement between two nodes:
\begin{equation}
P_1\left(L\right)=\mathcal{E}_{\mathrm{b}}\mathcal{E}_{\mathrm{m}}^{2}\mathcal{E}_{\mathrm{d}}^{2}e^{-\gamma L}.\label{eq:P-bsm-L}
\end{equation}
where $\mathcal{E}_{\mathrm{m}}$ and $\mathcal{E}_{\mathrm{d}}$ are memories and detectors efficiencies, correspondingly.  

Knowing the timestamp and an average number of attempts until successful entanglement generation, one can calculate how much time it takes for one successful entanglement generation.  Furthermore, entanglement can exist only for a limited time $\tau_{\mathrm{mem}}$, called memory lifetime.  If the entanglement degrades too fast before the photon reaches BSM, it would be impossible to establish a quantum link between the two nodes. In SeQUeNCe, memory lifetime is a fixed parameter counting lifetime of an entanglement since its inception when a qubit had successfully entangled with the photon. Once two photons with different scheduled expiration timestamps $t_{1}$ and $t_{2}$ are successfully measured by BSM, the lifetime for the generated entanglement between the two qubits is equal to  $\min\{t_{1},\,t_{2}\}$. Defining the entanglement generation rate $\mathcal{R}$ as reversed average time spent on the generation of entanglement, one can write the following expression:
\begin{equation}
\mathcal{R}=\left\{ \begin{aligned}\begin{aligned}\frac{v}{4L}\mathcal{E}_{\mathrm{b}}\mathcal{E}_{\mathrm{m}}^{2}\mathcal{E}_{\mathrm{d}}^{2}e^{-\gamma L}, & \;L<2v\tau_{\mathrm{mem}},\\
0, & \;L\geq 2v\tau_{\mathrm{mem}}.
\end{aligned}
\end{aligned}
\right.\label{eq:rate-no-reps}
\end{equation}
where the top line is given by the effective attempt frequency $v/4L$ times $P_1$ in Eq.(\ref{eq:P-bsm-L}). 
We assume that the repetition rate for the quantum state preparation is higher than  $v/L$ and it is not a limiting factor in determining  entanglement generation rate, which is not the case in the small $L$ limit~\cite{bestrate1,bestrate2}. In Fig.~\ref{fig:elementary}b, we find an excellent agreement between the analytical expression in Eq.~(\ref{eq:rate-no-reps}) and numerical simulations.

\begin{figure}[thp]
\centering{\includegraphics[width=0.99\columnwidth]{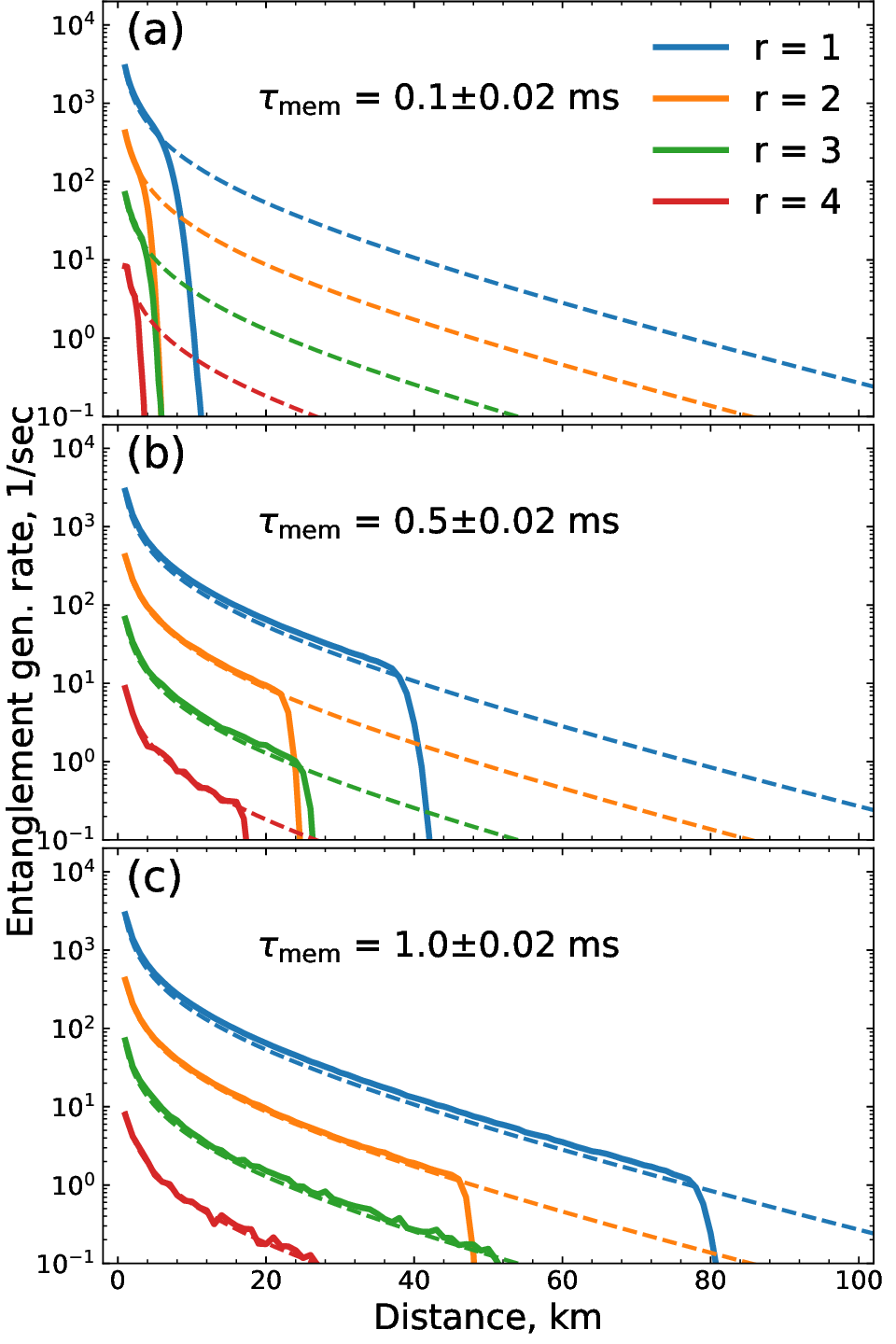} }
\caption{ {\it Synchronous generation,} entanglement generation rate versus distance between the end-nodes for a different number of repeaters and memory lifetimes: (a) $\tau_{mem}=0.1\pm0.02$ ms, (b) $\tau_{mem}=0.5\pm0.02$ ms, (c) $\tau_{mem}=1.0\pm0.02$ ms. Solid lines are simulations with SeQUeNCe, and dashed lines show the analytical model results using Eq.~\eqref{eq:reps-syn}. In the analytical model, we assume $\tau_{\mathrm{mem}}=\infty$.  The parameters of quantum memories, BSM, detectors, and optical channels are the same as in Fig.~\ref{fig:elementary}b: $\mathcal{E}_{\mathrm{m}}=90\%$, $\mathcal{E}_{\mathrm{s}}=50\%$, $\mathcal{E}_{\mathrm{d}}=80\%$, $\alpha=0.2$ dB/km and $v=2\times10^{5}$ km/s.}
\label{fig:rep-res-syn}
\end{figure}

\section{End-to-end quantum connection in a network}
Once an elementary quantum link between the nodes is established, it is possible to join an elementary link in a more extended end-to-end quantum link using the entanglement swapping protocol (see Fig.~\ref{fig:rep})~\cite{asadi2020protocols}. When establishing the end-to-end connection between some pairs of distant nodes, some algorithms would choose the optimal path from one end-node to another via intermediate nodes, considered quantum routers~\cite{Yuan2015,Schoute2016,Pant2019}. Once the path is determined, the problem is reduced to establishing connections between the successive nodes in the chain. Quantum routers give us dual benefits. The first one allows achieving connectivity by a much smaller number of physical communication links between the nodes than $N\left(N-1\right)/2$, where the $N$ is the number of nodes in the network. The second one is overcoming the exponential factor in~Eq.~\eqref{eq:rate-no-reps}. In the case of a network, $L$ in Eq.~\eqref{eq:rate-no-reps} is the communication path length between the end-nodes. To recover the entanglement/fidelity losses, we divide the distance between the end-nodes A and B into several segments by setting up a number $r$ of intermediate nodes (or quantum repeaters), as shown in Fig.~\ref{fig:rep} for the case of $r=3$.  For simplicity, we consider only a chain-like topology of the networks with equal distances between the adjacent nodes and generation end-to-end quantum connectivity between the end-nodes.

Suppose all the links among  adjacent qubits in Fig.~\ref{fig:rep} have been established by either synchronous or independent generation (discussed below in detail).  The next step is to perform entanglement swapping~\cite{asadi2020protocols} within each of the repeater nodes simultaneously, which would allow qubits in the end-nodes to become entangled if all the swappings are successful. Two kinds of procedures contribute to the time duration of this process. The first one is the time the middle nodes need to acknowledge the swapping results to the side nodes with entangled qubits. The second one is the time needed for all the nodes involved in the swapping to acknowledge the C-node about the operation results.  The total time spent on the entanglement swapping can be estimated as $\tilde{n}L/v$, where $\tilde{n}=\log_2\left(r+1\right)$ is the number of stages at which the total entanglement swapping procedure is done. We assume that the number of repeaters equals to $r=2^{\tilde{n}}-1$. In Fig.~\ref{fig:rep}, we demonstrate a case of $r=3$, which would require $\tilde{n}=2$ stages to establish entanglement between the end-nodes. Our numerical simulations show that analytical expressions for the entanglement generate rate in Eq.~(\ref{eq:reps-syn}) and (\ref{eq:reps-ind}), derived under the assumption that $r=2^{\tilde{n}}-1$,  works very well, even for the cases when $\tilde{n}$ is not an integer.

The probability of success of swapping operation is $\mathcal{E}_{\mathrm{s}}$.
If swapping fails at any elementary link, the whole
process fails, and all the links between the adjacent nodes must be
discarded and regenerated. When a particular swapping fails, the rest of the links are not as {\it fresh} as newly generated anymore. If the established links had to wait for the regeneration of the broken links, they would have degraded further and become less reliable. Thus, everything has to be done from scratch in this approach, even if a single link fails to establish. The whole process is repeated until entanglement between the end-nodes is established.

\section{Synchronous generation of entanglements between adjacent nodes}
Let us consider the case of entanglement generation with $r$ repeaters and $r+1$ elementary links. We try to generate entanglement between adjacent repeater nodes simultaneously and discard an attempt if one of the links fails. In such a scheme, we obtain the {\it freshest} entanglement because the entanglement between adjacent nodes is established at the same time, and there is no need for waiting. Due to the shorter lengths between neighboring repeater nodes, communication time between repeaters decreases, such that the total time needed for one generation attempt becomes $3L/\left(v\left(r+1\right)\right)+L/v$. This time now determines the effective attempt frequency, which was $v/4L$ in Eq.~(\ref{eq:rate-no-reps}) for a single link. The entanglement swapping is needed only when all adjacent links are successfully generated, which is a rare event: in the synchronous approach, it usually takes repeated attempts before all links are established simultaneously, while in the independent approach, there is usually a long wait before all the links are established. Therefore, we neglect the time needed for entanglement swapping  but take into account the time needed for the C-node (in the middle of the repeaters chain) to inform all the nodes about the start time of synchronous entanglement generation and the time needed for the nodes to return their elementary link generation status to the C-node. The probability of successful generation of all the $r+1$ adjacent links simultaneously equals to $\left[P_1\left(L/\left[r+1\right]\right)\right]^{r+1}$. The probability of successful swapping of $r$ links is given by  $\mathcal{E}_{\mathrm{s}}^{r}$ (see Fig.~\ref{fig:rep} for the case of $r=3$). A product of those two probabilities times the effective attempt frequency leads to the following entanglement generation rate:
\begin{equation}
\mathcal{R}_{\mathrm{syn}}=
\frac{v}{L}
\left[\frac{3}{r+1}+1\right]^{-1}
\mathcal{E}_{\mathrm{b}}^{r+1}\mathcal{E}_{\mathrm{s}}^{r}
(\mathcal{E}_{\mathrm{m}}\mathcal{E}_{\mathrm{d}})^{2\left(r+1\right)}
e^{-\gamma L}.\label{eq:reps-syn}
\end{equation}
Note that Eq.~(\ref{eq:reps-syn}) reduces to Eq.~(\ref{eq:rate-no-reps}) in the limit of no repeaters, i.e., $r=0$. 

\begin{figure}[htp]
\centering{\includegraphics[width=0.99\columnwidth]{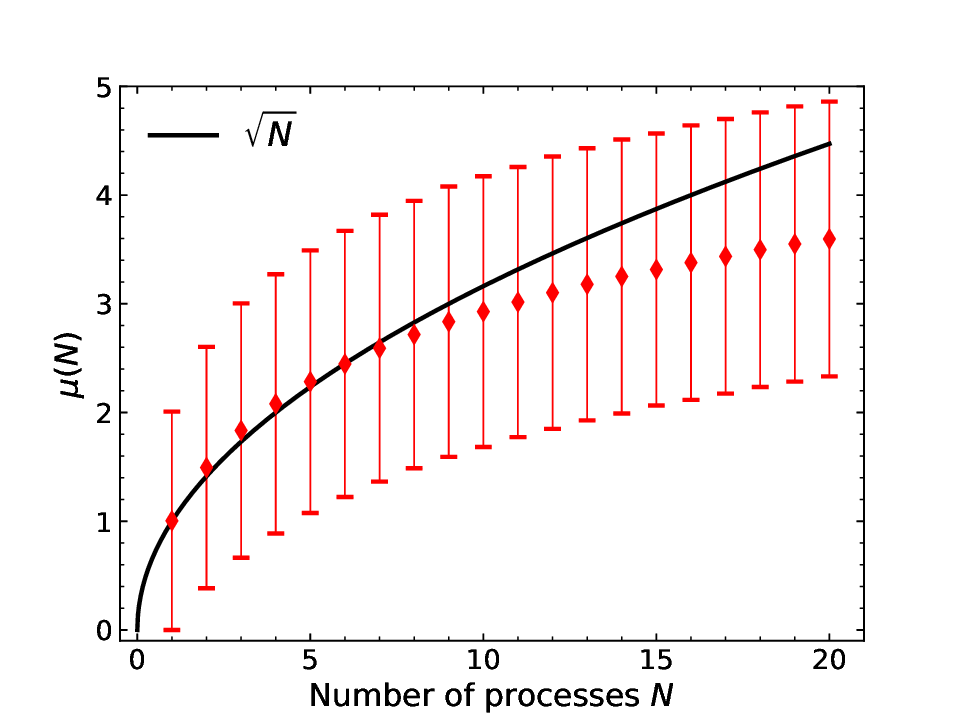}}
\caption{Simulations of the maximum number of attempts (see the highest blue column in panel (a)) until success in all elementary links, normalized to the statistical average for each link, as a function of the number of elementary links $N$. The distribution in  Eq.~$\eqref{eq:P-k-bsm}$ describes each elementary link with $P_1=10^{-3}$. Error bars show a standard deviation of $\mu(N)$. Simulations used $10^6$ repetitions.}
\label{fig:num-tries}
\end{figure}

For simplicity, when analyzing the cases with an arbitrary number of repeaters, we do not account for the memory coherence time in analytical expression Eq.~(\ref{eq:reps-syn}). However, one should keep in mind that despite its efficiency in saving  {\it freshness} of the generated entanglements, the described scheme cannot overcome the limit imposed by the distance between the end-nodes, i.e., $L\lesssim0.5v\tau_{\mathrm{mem}}$. Otherwise, the entangled memories would degrade during the photon travel time of about $2L/v/(r+1)+\tilde{n}L/v$, as in the no-repeater case.

\begin{figure}[bhp]
\centering{\includegraphics[width=0.99\columnwidth]{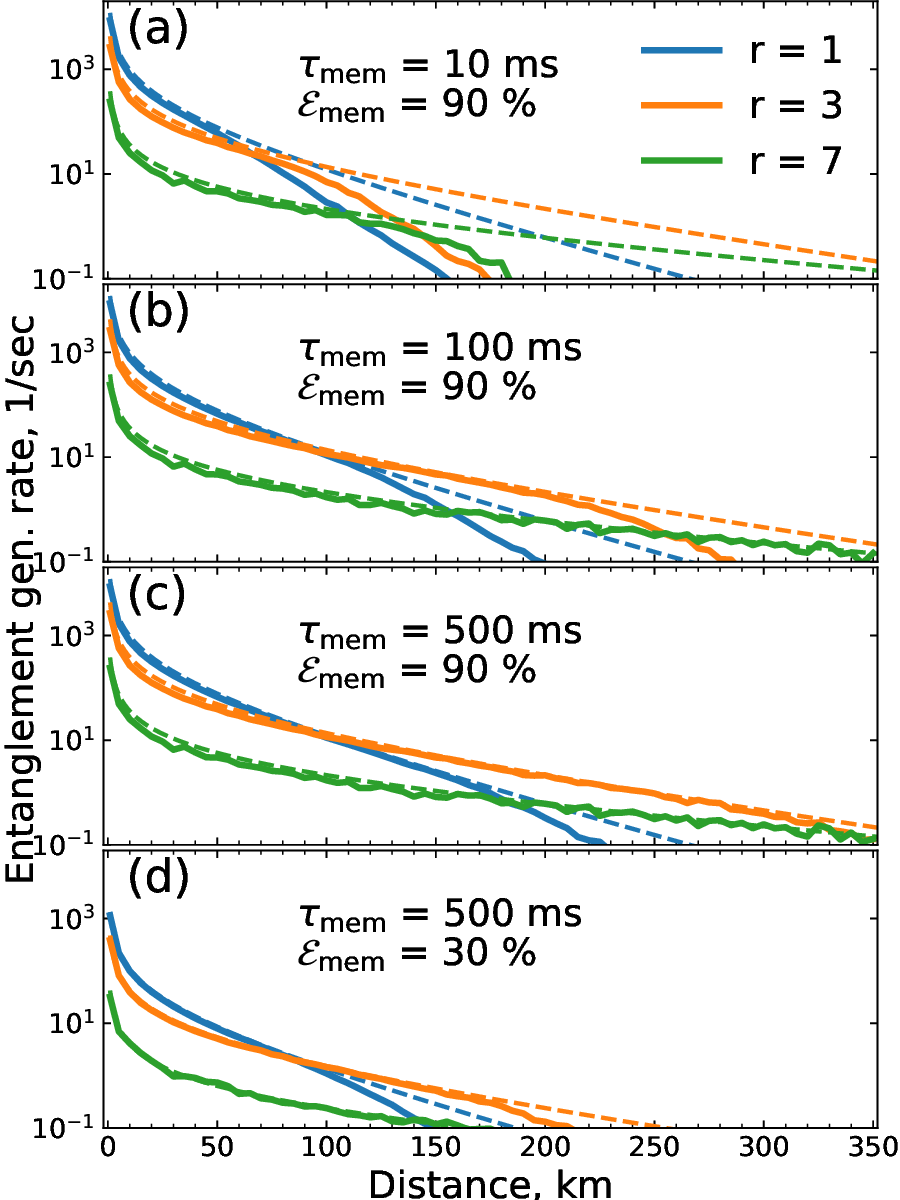} }
\caption{ {\it Independent generation,} entanglement generation rate versus distance between the end-nodes for a different number of repeaters and memory lifetimes:  (a) $\tau_{mem}=10$ ms, (b) $\tau_{mem}=100$ ms, (c) and (d) $\tau_{mem}=500$ ms.
Solid lines are simulations with SeQUeNCe, and dashed lines show the analytical model results using Eq.~\eqref{eq:reps-ind}. In the analytical model, we assume $\tau_{\mathrm{mem}}=\infty$.  The parameters of quantum memories, BSM, detectors, and optical channels are the same as in Fig.~\ref{fig:rep-res-syn} except for panel (d), where we used a smaller memory efficiency $\mathcal{E}_{\mathrm{m}}=30$\%.}
\label{fig:rep-res-ind}
\end{figure}
\noindent

Eq.~\eqref{eq:reps-syn}  was derived in the limit when the probability of photon loss in a quantum channel is very low. In this case, the most probable outcome of the generation attempt is a failure, and  the outcomes whose duration is longer than $3L/(v(r+1))+L/v$ are negligible. However, plots in Fig.~\ref{fig:rep-res-syn} show that this assumption is also applicable in the cases of relatively short distances between the end-nodes, i.e., $L \simeq 1$ km. The minor discrepancies between Eq.~\eqref{eq:reps-syn}  and the numerical simulations are caused by the lack of path optimization notifying signal sent to the C-node from the nodes generating elementary links. In other words, we assume that the notification is sent only by the node that is closer to the C-node. Thus, the time delay between the Barrett-Kok protocol operation while generating the elementary links instead of $L/v$ should be smaller by at least $L/v/(r+1)$. When we simulate cases when $r+1\neq2^l$, $l\in\mathcal{N}$,  the C-node is not exactly in the middle of the chain. A more precise version of Eq.~\eqref{eq:reps-syn}  is significantly more complicated. Thus we keep the current form, which gives the single swapping round duration  $0.5L/v$  as in the case $r+1=2^l$, $l\in\mathcal{N}$.  In the current implementation of our simulations, we add $L/v$  delay between plain SeQUeNCe’s swapping protocol operation rounds to account for the delays due to the C-node notifying nodes about all the elementary links generation attempts.

As shown in Eq.~(\ref{eq:reps-syn}), one feature of synchronous entanglement generation is that it gives an even lower entanglement generation rate than in the no-repeater case. However, such a scheme allows to utilize the intermediate nodes as routers and organize multiple nodes into a quantum network without the need to lay communication fiber between each pair of the network nodes. In addition, the lifetime of quantum memory needs to be much smaller than in the case of the independent generation scheme discussed below. This can be essential if solid-state-based scalable hardware components are employed for quantum repeaters.

\section{Independent generation of entanglements between adjacent nodes}
Another way to generate entanglement links between the end-nodes involves generation links among adjacent nodes independently, without discarding already established links. In this scheme, a link between the end-nodes can be generated with a much higher probability than in the synchronous protocol. Consequently, some of the links established early on would have to wait for a longer time, making decoherence a much more stringent requirement than in the case of the synchronous protocol. 

In this scenario, the tyranny of fiber losses is overcome since the probability for establishing $r+1$ links is no longer $P_1^{r+1}$,  but instead, it is proportional to $P_1$. Our numerical simulations aim to establish the coefficient of proportionally.  We show that the longer the time difference between the completion of the first and the last links, the more probabilistic events are tested before the success. As a result, a better statistic is obtained, which is the case for any Monte-Carlo type simulation. Since the success probability for the generation of an elementary link $P_1$ is given by Eq.~\eqref{eq:P-bsm-L}, the success probability after $k$ attempts is given by:
\begin{equation}
P_{k}=\left(1-P_1\right)^{k-1}P_1.\label{eq:P-k-bsm}
\end{equation}
Note that probability distribution $P_k$ is normalized, since $\sum_{k=1}^{\infty}P_k=1$. 
The average number of attempts till success and its standard deviation are given by:
\begin{eqnarray}
&&\overline{k}=\sum_{k=1}^{\infty}kP_k=-P_1\frac{d}{dP_1}\sum_{k=1}^{\infty} (1-P_1)^k=\frac{1}{P_1}
\nonumber \\
&&\overline{k^2}+\overline{k}=\sum_{k=1}^{\infty}k(k+1)P_k=P_1\frac{d^2}{dP_1^2}\sum_{k=1}^{\infty} (1-P_1)^{k+1}=\frac{2}{P^2_1}
\nonumber \\
&&\quad\sqrt{\overline{\delta k^{2}}}=\sqrt{\overline{k^{2}}-\overline{k}^2}=\sqrt{\frac{1}{P_1^{2}}-\frac{1}{P_1}},\label{eq:P-k-mean-stdev}
\end{eqnarray}
where $\overline{f_{k}}$ means averaging the function $f_{k}$ over
the probability distribution given by Eq.~(\ref{eq:P-k-bsm}). In most cases,$P_1\ll 1$, so that $\sqrt{\overline{\delta k^{2}}}\approx1/P_1$.

We simulate a set of $N$ processes corresponding to $N=r+1$ elementary links in the chain with $r$ repeaters. Each process is described by the distribution in Eq.~(\ref{eq:P-k-bsm}). Using pseudo-random number generation, we find for each process a number of attempts $k_i$ till success, where $i=1,\dotsc,N$. We define a value $\mu(N)$ as a maximum number of attempts normalized to the statistical average number $\overline{k}$, such that $\mu(N)=\max\{k_i\}/\overline{k}$. Fig.~\ref{fig:num-tries} shows the result for $\mu(N)$  using elementary link success probability $P_1=0.001$. For smaller values of $P_1$, the result does not depend on $P_1$.

For $N\leq8$, the obtained dependence can be approximated well by $\mu\left(N\right)\approx\sqrt{N}$ dependence. Therefore, in the case when $r+1$ entanglements are generated by the pairs of nodes independently, the average number of attempts till the last link is generated is given by $\mu\left(r+1\right)/P_1$. Analogously to Eqs.~$\eqref{eq:rate-no-reps}$ and (\ref{eq:reps-syn}), one can obtain an entanglement generation rate as a product of probability to generate $N$ independent links, $P_1\left(L/\left[r+1\right]\right)/\mu\left(r+1\right)$, entanglement swapping probability $\mathcal{E}_{\mathrm{s}}^{r}$, and effective attempt frequency to arrive at:  
\begin{equation}
\mathcal{R}_{\mathrm{ind}}=\left[\frac{3\mu\left(r+1\right)}{r+1}\right]^{-1}\frac{v}{L}\mathcal{E}_{\mathrm{b}}\mathcal{E}_{\mathrm{s}}^{r}\mathcal{E}_{\mathrm{m}}^{2}\mathcal{E}_{\mathrm{d}}^{2}e^{-\gamma L/\left(r+1\right)}.\label{eq:reps-ind}
\end{equation}
Note that the effective attempt frequency, in this case, is reduced as compared to the synchronous case in Eq.~(\ref{eq:reps-syn}) because the time $L/v$ needed to inform the control node that all independent links are established is much shorter than the time needed to establish the links.

The average age of the oldest entangled memory, when the end-to-end entanglement is established, equals to:
\begin{equation}
\varDelta t=\mathcal{R}_{\mathrm{ind}}^{-1}+\frac{L}{v}\log_2\left(r+1\right).
\label{eq:dt}
\end{equation}
where the first term in Eq.~(\ref{eq:dt}) corresponds to the time needed to generate all adjacent  links and the second term for entanglement swapping. 

That time is much longer than in the synchronous generation scenario, and from the previous two equations, one finds that $\varDelta t\gg L/v$. Consequently, quantum memory lifetime should be at least $\tau_{\mathrm{mem}}\gtrsim\varDelta t$ to maintain quantum states during the entanglement generation time between the end-nodes. However, the simulations in Fig.~\ref{fig:rep-res-ind} show that indeed after $\mathcal{R}_{\mathrm{ind}}^{-1}$  reaches approximately $\tau_\mathrm{mem}^{-1}$, the entanglement generation rate starts to degrade faster than the analytical expression Eq.~\eqref{eq:reps-ind}, although not as rapidly as in the synchronous scheme considered above.

Fig.~\ref{fig:rep-res-ind} demonstrates that in the limit of long memory lifetime, Eq.~\eqref{eq:reps-ind} works very well for predicting entanglement generation rate as a function of distance, a number of repeaters, and parameters defining hardware performance. Therefore, it can be used for a reverse problem of finding the required parameters of the quantum local area network hardware components for a given distance and entanglement generation rate.

\section{Conclusion}

We have simulated two types of protocols of elementary quantum links in a chain-like network supporting quantum entanglement between the end-nodes. The hardware components and basic protocols for entanglement generation and swapping are adopted from the SeQUeNCe package. We have explored the effects of finite memory lifetime on entanglement generation in a quantum network for two entanglement swapping protocols. For the synchronous generation of entanglement between adjacent nodes, the advantages of quantum repeaters are limited. On the other hand, for the independent generation of entanglement between adjacent nodes, additional repeaters enable communication at much longer distances between the end-nodes. We present analytical solutions for the entanglement generation rate for both scenarios, which are almost exact in the limit of infinitely long quantum memory lifetimes. In both cases, our numerical simulations demonstrate that entanglement generation degrades due to a finite quantum memory lifetime, whereas this degradation is less severe in the independent entanglement generation scenario. Our results demonstrate the ultimate performance of a quantum network as a function of the parameters defined by the network's hardware components, the number of repeaters, the distance between the end-nodes, and the lifetime of the quantum memories.

\section*{Acknowledgments}

We gratefully acknowledge Kathy-Anne Soderberg and David Hucul from AFRL, Rome, NY, for insightful discussions and funding from the Vice President for Research and Economic Development (VPRED) at the University at Buffalo and SUNY Research Seed Grant Program, and computational facilities at the Center for Computational Research at the University at Buffalo (\begin{sloppy}\url{http://hdl.handle.net/10477/79221}\end{sloppy}). X. H. acknowledges support by US ARO via grant W911NF1710257. E.F. acknowledges support by NSF via grant NSF PHY-1707919. 

\section*{Conflict of interest}

The authors have no conflicts to disclose.


\section*{Data availability}
The data that supports the findings of this study are available within the article.



\section*{References}
\bibliography{references}
\end{document}